# On the kinetics of homogeneous nucleation of incoherent precipitates in solid solutions

M.S. Veshchunov[*)]

Nuclear Safety Institute (IBRAE), Russian Academy of Sciences,

52, B. Tulskaya, Moscow 115191, Russian Federation

**Abstract**

By critically examining the traditional theory of homogeneous nucleation of precipitates in solid solutions, it is revealed that the theory's assertion regarding an increase in the nucleation free energy due to elastic strain associated with the difference in atomic volumes between the two phases is applicable to coherent precipitates, but becomes incorrect when applied to incoherent precipitates. This conclusion is obtained by accounting for thermal point defects in the matrix, which can be absorbed at the interface between an incoherent particle and the matrix during nucleation, thereby relieving elastic stresses. Accordingly, a new kinetic model based on the Reiss theory for binary nucleation is proposed for predicting the nucleation rate of incoherent precipitates by agglomeration of solute atoms and point defects, with a further extension to account for excess vacancies formed under non-equilibrium conditions of quenching experiments.

[*)] Corresponding author. E-mail: msvesh@gmail.com

## 1. Introduction

Precipitation in solid materials often results in the generation of misfit elastic strain energy, arising from volume and/or shape incompatibility between the new phase nuclei and the parent phase matrix. These changes encounter resistance due to the stiffness of the surrounding matrix, inducing elastic strains. Both the elastic strain energy and the surface energy produced by the new phase nucleus make a positive contribution to the free energy of its formation, thereby serving as a barrier to nucleation. The magnitude of the elastic energy term typically depends on factors such as the shape of the cluster, the mismatch between the cluster and matrix, and whether the interface between them is coherent, semicoherent, or incoherent [1–3].

The surface energy of coherent interfaces is usually several times less than the surface energy of an incoherent interface [3]. Consequently, homogeneous nucleation is usually observed in cases where the nucleus interface is coherent and the interfacial energy is relatively low.

However, the significant variance in atomic volumes and crystal structures can lead to a predominantly incoherent interface between the precipitate and matrix. For example, there is no indication of coherent clustering of silicon in fcc aluminium alloys. Silicon possesses a diamond cubic structure, and there is no evidence supporting the existence of a low-energy face-centred cubic modification [4]. Additionally, silicon has a larger atomic volume than aluminium, resulting in a deformation strain of $\approx 0.2$, which may be too large to form a coherent interface. Homogeneous precipitation of silicon from a supersaturated solid solution in aluminium typically occurs in the form of equiaxed particles randomly dispersed within the grains, as confirmed by X-ray diffraction, light microscopy, and electron microscopy analyses [5, 6]. Similar observations using transmission electron microscopy (TEM) have been reported in studies of Al-Si alloys following various quenching and pre-aging treatments [7], and more recently using TEM, dilatometry and differential scanning calorimetry [8].

Contemporary concepts of strengthening, fracture and precipitation in alloys propose that ultrafine dispersoids in an fcc Al-matrix provide the optimal microstructure for achieving ultra-high strength and fracture resistance. This microstructure is primarily found in two binary alloy systems: Al-Si and Al-Ge. Consequently, these alloys not only serve as models for investigating the mechanisms behind the nucleation of incoherent phases but also lay the basis for a novel category of technical aluminium alloys [9].

In the traditional nucleation theory for a transformation where strain energy effects are not negligible, the formation of a nucleus of given size will require an increase in the Gibbs free energy of the nucleus formation, $\Delta G_0$, with the addition of the elastic energy term (see, e.g. [1–3]). The elastic energy is proportional to the number $x$ of atoms in the nucleus, so that

$$\Delta G_0(x) = x(-\Delta g_b + \Delta g_{el}) + \Delta G_{surf}, \tag{1}$$

where $\Delta g_b = kT \ln S_x$ is the bulk free energy change per atom involved in the formation of the nucleus in the solid solution with the supersaturation $S_x$; $\Delta g_{el}$ is the elastic energy per atom; $\Delta G_{surf} = \gamma 4\pi R^2$ is the surface energy of the nucleus with surface tension $\gamma$ (which for incoherent inclusions has a characteristic value of ~ 1 J·m$^{-2}$ [3]).

For an incoherent spherical nucleus of radius $R_x = (3\Omega/4\pi)^{1/3} x^{1/3}$, the misfitting sphere model of Nabarro [10] may be applied to give the total strain energy (cf. [1]),

$$\Delta G_{el} = x\Delta g_{el} = 6\mu \left(\frac{3K}{3K+4\mu}\right) \delta^2 V_p, \tag{2}$$

where $\delta \approx (V_p - V_m)/3V_p \ll 1$ is the transformation strain due to forming the particle of a volume $V_p = x\Omega$ placed in a spherical cavity of a volume $V_m = x\Omega_m$; $\Omega$ and $\Omega_m$ are the atomic volumes in the particle and in the matrix respectively; $\mu \approx 25$ MPa is the shear modulus of the Al matrix, and $K \approx 100$ MPa is the bulk modulus of the Si particle. In the limit $3K \gg 4\mu$, this equation can be simplified,

$$\Delta G_{el} = x\Delta g_{el} \approx 6\mu\delta^2 V_p, \quad (2')$$

as is often assumed in the literature (e.g. in [2]). In the following analysis, the traditional Eq. (2') will be used with a renormalized value of the shear modulus of the Al matrix, $\mu \to \mu\left(\frac{3K}{3K+4\mu}\right) \approx \frac{3}{4}\mu \approx$ 19 MPa, which correctly transforms Eq. (2') to a more accurate Eq. (2).

Minimization of Eq. (1), $\partial \Delta G_0(x)/\partial x = 0$, gives the critical nucleus size,

$$x^* = \frac{32\pi}{3}\left(\frac{\gamma}{kT \ln S_x - \Delta g_{el}}\right)^3 \Omega^2 = \frac{32\pi}{3}\left(\frac{\gamma}{kT \ln S_x - 6\mu\Omega\delta^2}\right)^3 \Omega^2, \quad (3)$$

and the formation free energy of the critical nucleus,

$$\Delta G_0(x^*) = \frac{16\pi}{3}\frac{\gamma^3 \Omega^2}{(kT \ln S_x - 6\mu\Omega\delta^2)^2}, \quad (4)$$

which demonstrates that the misfit strain reduces the effective driving force for precipitation and increases the critical supersaturation to $S_x^* = \exp(6\mu\Omega\delta^2/kT) > 1$ [1–3].

In this article, based on a critical examination of the traditional theory of the nucleation of incoherent precipitates in solid solutions based on Eqs (3) and (4), it will be shown that accounting for thermal point defects in the parent phase, which can be absorbed at the incoherent particle-matrix interface during particle nucleation, eliminates the nucleation barrier associated with the elastic strain energy generated by the nuclei of the new phase. Building upon this insight, a new kinetic model for the nucleation of incoherent precipitates will be formulated within the framework of the Reiss theory for binary homogeneous nucleation, specifically tailored to a binary system comprising an oversaturated solid solution and thermal vacancies. Furthermore, the kinetic model proposed herein can be extended to incorporate excess vacancies generated under non-equilibrium conditions, such as during quenching, which can significantly improve the predictions of Russel's model [11], developed within the framework of traditional nucleation theory for single component (unary) systems.

## 2. Critical analysis of the traditional theory

It is important to highlight that Eqs (3) and (4) hold true specifically for coherent particles. This is because vacancies and self-interstitials can only be trapped (or *ad*sorbed) at the particle-matrix interface, thereby leaving the elastic energy of the particle unaffected.

However, in the case of incoherent particles, where point defects are *ab*sorbed at the matrix interface with the particle, a noticeable change in the phase transformation mechanism can occur. With the absorption of vacancies and the emission of self-interstitials (with a total number of $n$), the interface of an oversized particle moves outward, resulting in a simultaneous expansion of the cavity volume (where the particle is inserted), $V_m \to V_m' = (x + n)\Omega_m$, the radius of the interface, $R \to R' = (3\Omega_m/4\pi)^{1/3}(x + n)^{1/3} = [3\Omega/4\pi(1+\varphi)]^{1/3}(x + n)^{1/3}$ and its surface area, $S \to S'$, where $\varphi = (\Omega - \Omega_m)/\Omega_m$. In turn, this leads to a decrease in the elastic energy $\Delta G_{el}$ (due to a decrease in the transformation strain $\delta \to \delta' = (V_p - V_m')/3V_p = \frac{1}{3}\left(\varphi - \frac{n}{x}\right)\left(\frac{1}{1+\varphi}\right)$) and an increase in the surface energy $\Delta G_{surf}$.

Since chemical potential of equilibrium point defects in the matrix is zero, the free energy, Eq. (1), does not change due to their absorption (or emission) at the interface and therefore takes the form

$$\Delta G_0(x,n) = -kTx \ln S_x + 4\pi\gamma \left(\frac{3}{4\pi}\Omega_m\right)^{\frac{2}{3}} (x+n)^{\frac{2}{3}} + \frac{2\mu\Omega}{3}\left(\frac{1}{1+\varphi}\right)^2 x \left(\varphi - \frac{n}{x}\right)^2. \quad (5)$$

Minimization of Eq. (5) with respect to the two variables, $\partial \Delta G_0(x,n)/\partial x = \partial \Delta G_0(x,n)/\partial n = 0$, gives the critical nucleus size,

$$x^* = \frac{288\pi\gamma^3\Omega^2}{[4\mu\Omega(1-q)]^3 q} \approx \frac{32\pi}{3}\left(\frac{\gamma}{kT\ln S_x}\right)^3 \Omega^2 \left(1 - \frac{3kT}{8\mu\Omega}\ln S_x\right) \approx \frac{32\pi}{3}\left(\frac{\gamma}{kT\ln S_x}\right)^3 \Omega^2, \quad (6)$$

where $q = \left(1 - \frac{3kT}{2\mu\Omega}\ln S_x\right)^{\frac{1}{2}}$, or $q \approx 1 - \frac{3kT}{4\mu\Omega}\ln S_x$ in the first approximation in a small parameter $3kT/4\mu\Omega \approx 10^{-2} \ll 1$ (at test temperature $\approx 300$ K), and

$$\frac{n^*}{x^*} = (\varphi + 1)q - 1 \approx \varphi - \frac{3kT}{4\mu\Omega}(1+\varphi)\ln S_x, \quad (7)$$

whereas the formation free energy of the critical nucleus is calculated as

$$\Delta G_0^* \equiv \Delta G_0(x^*, n^*) = \frac{48\pi\gamma^3\Omega^2[4\mu\Omega(q-1)+6kT\ln S_x]}{[4\mu\Omega(1-q)]^3 q} \approx \frac{16\pi}{3}\left(\frac{1}{kT\ln S_x}\right)^2 \gamma^3\Omega^2 \left[1 + \frac{3}{4}\frac{kT}{\mu\Omega}\ln S_x\right] \approx \frac{16\pi}{3}\frac{\gamma^3\Omega^2}{(kT\ln S_x)^2}, \quad (8)$$

which demonstrates that the contribution of the elastic energy to the nucleation barrier becomes negligible, and, consequently, Eqs (3) and (4) are invalid for incoherent precipitates. Therefore, the traditional theory's assertion that the strain energy caused by the difference in atomic volumes of the two phases increases the barrier to nucleation is inaccurate and requires revision to assess the nucleation rate of incoherent particles using Eq. (8).

It is important to note from Eq. (7) that for a relatively small misfit strain of oversized particles with $0 < \varphi < \frac{3kT}{4\mu\Omega}\ln S_x$, the critical nucleus does not absorb but emits vacancies, $n^* < 0$; this occurs because, at such misfits, the decrease in the nucleus surface energy due to the emission of vacancies prevails over the increase in the elastic strain energy. This demonstrates the importance of self-consistent consideration of changes in the nucleus volume and surface due to absorption/emission of point defects.

In typical experiments with supersaturated alloys (see, e.g. [5–7]), the maximum number of particles was nucleated when the samples were quenched, creating an excess of vacancies in the matrix (with a supersaturation level $S_v = c_v/c_v^{(0)}$, where $c_v$ is the dimensionless non-equilibrium concentration of vacancies and $c_v^{(0)}$ is its thermal value), which can significantly accelerate the nucleation process. More generally, an excess of self-interstitials in the quenched (non-equilibrium) matrix (with supersaturation $S_i = c_i/c_i^{(0)}$), should be additionally taken into account. However, the condition $c_i(T_H)/c_v(T_H) = c_i^{(0)}/c_v^{(0)} \ll 1$ in the equilibrium crystal at the homogenation temperature $T_H$ before quenching, which is normally realised in metals (since self-interstitials have rather high formation enthalpies compared to vacancies [12]), leads to the survival of only excess vacancies due to fast annihilation of point defects during subsequent cooling.

The influence of excess vacancies on the nucleation barrier was considered by Russel [11]. However, in the absence of excess vacancies, his model can be reduced to the above Eq. (5) with some modifications. Namely, the free energy of formation of an incoherent particle took into account the change in its volume due to the absorption of vacancies, but neglected the increase in the interface area (which introduces some inconsistency in the model predictions for small $\varphi$, as explained above).

A more consistent result can be obtained by generalizing Eq. (5) to take into account excess vacancies formed under non-equilibrium (quenching) conditions, which makes it possible to refine Russel's model. For this, an additional term, $-kTn \ln S_v$, describing the variation of the free energy of $n$ vacancies due to absorption at the interface, has to be implemented in Eq. (5), leading to

$$\Delta G_0(x,n) = -kTx \ln S_x - kTn \ln S_v + 4\pi\gamma \left(\frac{3}{4\pi}\Omega_m\right)^{\frac{2}{3}} (x+n)^{\frac{2}{3}} + \frac{2}{3}\mu\Omega \left(\frac{1}{1+\varphi}\right)^2 x \left(\varphi - \frac{n}{x}\right)^2, \quad (9)$$

which minimization with respect to $x$ and $n$ in the critical point gives in the first approximation in $3kT/4\mu\Omega \ll 1$,

$$\frac{n^*}{x^*} = (\varphi+1)\tilde{q} - 1 \approx \varphi - (\varphi+1)\frac{3kT}{4\mu\Omega} \ln \frac{S_x}{S_v}, \quad (10)$$

where $\tilde{q} = \left(1 - \frac{3kT}{2\mu\Omega} \ln \frac{S_x}{S_v}\right)^{1/2}$, and

$$x^* = \frac{288\pi\gamma^3\Omega^2}{[3kT(1+\varphi)\ln S_v + 4\mu\Omega(1-\tilde{q})]^3 \tilde{q}} \approx \frac{32\pi}{3}\left(\frac{\gamma}{kT}\right)^3 \frac{\Omega^2}{\left[\ln S_x + \varphi \ln S_v + \frac{3kT}{8\mu\Omega}\left(\ln\frac{S_x}{S_v}\right)^2\right]^3} \left(1 + \frac{3kT}{4\mu\Omega}\ln\frac{S_x}{S_v}\right). \quad (11)$$

Correspondingly, the expression for the formation free energy of the critical nucleus takes the form

$$\Delta G_0^* = \frac{48\pi\gamma^3\Omega^2[4\mu\Omega(q-1)+6kT\ln S_x+3kT[-2+(1+\varphi)\tilde{q}]]}{[4\mu\Omega(1-\tilde{q})+3kT(1+\varphi)\ln S_v]^3 \tilde{q}} \approx \frac{16\pi}{3}\frac{\gamma^3\Omega^2}{(kT)^2} \frac{1}{\left[\ln S_x + \varphi \ln S_v + \frac{3kT}{8\mu\Omega}\left(\ln\frac{S_x}{S_v}\right)^2\right]^3} \left[\ln S_x + \varphi \ln S_v + \frac{3kT}{2\mu\Omega}\ln\frac{S_x}{S_v}\ln S_v + \frac{15}{8}\frac{kT}{\mu\Omega}\left(\ln\frac{S_x}{S_v}\right)^2\right], \quad (12)$$

with the critical supersaturation $\ln S_x^* \approx -\varphi \ln S_v$.

This result demonstrates the effect of nucleation in undersaturated solutions (with $S_x < 1$) under non-equilibrium (quenching) conditions, first identified by Russel [11], who evaluated the nucleation barrier as

$$\Delta G_0^* = \frac{16\pi}{3}\frac{\gamma^3\Omega^2}{(kT)^2} \frac{1}{\left[\ln S_x + \varphi \ln S_v + \frac{3kT}{8\mu\Omega}\left(\ln\frac{S_x}{S_v}\right)^2\right]^2}, \quad (12')$$

which is significantly underestimated as compared to Eq. (12) in a small vicinity of the critical supersaturation (when the denominator tends to 0), but can be used in practical applications at higher supersaturations (when the critical size becomes small enough that the onset of nucleation can be observed).

However, to calculate the nucleation rate in the binary system of solute atoms and vacancies, Russel applied the traditional nucleation theory developed for single-component systems, which can change the pre-exponential factor of the nucleation rate by several orders of magnitude (as shown in the next Section).

### 3. Binary nucleation

According the analysis presented above, the problem of nucleation of equiaxed incoherent precipitates (observed in [5, 6]) is an example of homogeneous nucleation in binary systems, where the nucleus can be considered as a spherical particle of a new phase formed by agglomeration of solute atoms and vacancies. However, classical nucleation theory [13–15] (e.g. used by Russel [11]) was developed mainly in relation to single component (unary) systems and can lead to incorrect prediction of the pre-exponential kinetic factor in the nucleation rate.

This theory was generalized to the kinetics of nucleation in binary mixtures by Reiss [16]. In his theory, the parent phase is thought of as a mixture of molecules (monomers) of two components $X$ and $Y$ with number densities $N_x$ and $N_y$ (corresponding to dimensionless concentrations $c_x$ and $c_y$), respectively, together with clusters of all sizes and compositions. A particular molecular cluster is characterized by the numbers of single molecules (or monomers) $x$ and $y$ of species $X$ and $Y$, respectively, that it contains. Reiss showed that the critical point of unstable equilibrium corresponds in this case to a saddle point $(x^*, y^*)$ on the free energy surface $\Delta G_0(x, y)$. He characterized the rate of the transition by a two-dimensional flux vector $J(x, y)$ in the phase space of cluster sizes $x$, $y$, which is pronounced in the direction of the steepest descent of the free energy surface (the axis of the pass $x'$) that, in comparison with it, any lateral flow (in the perpendicular direction $y'$) may be neglected, i.e. $J_{y'} \approx 0$. Due to the steady state condition, $\text{div} J = \frac{\partial J_{x'}}{\partial x'} + \frac{\partial J_{y'}}{\partial y'} \approx \frac{\partial J_{x'}}{\partial x'} = 0$, this leads to $J_{x'} \approx J(y')$, which was calculated by Reiss as

$$J(y' - y^*) = f_0(x^*, y^*) \frac{\beta_x^* \beta_y^* (1+\tan^2\theta)}{\beta_y^* + \beta_x^* \tan^2\theta} \left(\frac{|D'_{11}|}{\pi kT}\right)^{1/2} \exp\left[-\frac{|\det \mathbf{D}|(y'-y^*)^2}{kT|D'_{11}|}\right], \tag{13}$$

where $f_0(x, y)$ is the equilibrium size distribution function,

$$f_0(x, y) = F \exp[-\Delta G_0(x, y)/kT], \tag{14}$$

$F$ is the so-called number density of potential nucleation sites, discussed below in Section 3.1; $\theta$ is the angle between the original axis $x$ and the axis of the pass $x'$; $\beta_i^* = \beta_i(x^*, y^*) = 4\pi D_i c_i R^* \Omega^{-1}$, $i = x, y$, are the arrival rates of monomers $X$ and $Y$ to the critical cluster $(x^*, y^*)$ of radius $R^*$; $D_{ij} = \frac{1}{2}\frac{\partial^2 \Delta G_0(x,y)}{\partial x_i \partial x_j}\Big|_{x^*, y^*}$ are elements of the matrix $\mathbf{D} = (D_{ij})$, which determinant is negative (in accordance with the properties of the saddle point, cf. [17]), $\det \mathbf{D} = D_{11}D_{22} - D_{12}^2 < 0$;

$$D'_{11} = \frac{1}{2}\frac{\partial^2 \Delta G_0(x', y')}{\partial x'^2}\Big|_{x^*, y^*} = D_{11}\cos^2\theta + D_{22}\sin^2\theta + 2D_{12}\sin\theta\cos\theta, \tag{15}$$

is the second derivative of $\Delta G_0$ in the direction $x'$ of the orthogonal coordinate system $(x', y')$ obtained by rotating the original coordinate system $(x, y)$ through the angle $\theta$; this derivative should be negative, $D'_{11} < 0$, to provide a maximum of the free energy at the critical point in the direction of the $x'$-axis.

Consequently, the nucleation rate, defined as the total flux of clusters through the critical zone,

$$\dot{N} = \int_{-\infty}^{\infty} J(y' - y^*) dy', \tag{16}$$

was calculated by Reiss as

$$\dot{N} \approx -f_0(x^*, y^*) \frac{\beta_x^* \beta_y^* (1+\tan^2\theta)}{\beta_y^* + \beta_x^* \tan^2\theta} D'_{11} \left(\frac{1}{D_{12}^2 - D_{11}D_{22}}\right)^{1/2}. \tag{17}$$

Reiss' theory was modified by Langer [18] (with subsequent reiteration by Stauffer [19]), who corrected the orientation of the flux vector in the direction parallel to the direction of the unstable mode at the saddle point (the new axis of the pass $x'$). The modified value of $\theta$ was explicitly calculated in [19] and later refined in [20] as

$$\tan\theta = s + (r + s^2)^{1/2}, \quad \text{if } D_{21} < 0, \tag{18}$$

and

$$\tan\theta = s - (r + s^2)^{1/2}, \quad \text{if } D_{21} > 0, \tag{19}$$

where $r = \beta_y^*/\beta_x^*$, $s = (d_a - rd_b)/2$, $d_a = -D_{11}/D_{12}$ and $d_b = -D_{22}/D_{12}$.

### 3.1. Number density of potential nucleation sites F

In the Reiss theory, given the total number density $N_{xy}$ of spherical clusters $X_xY_y$ consisting of $(x, y)$ monomers is small compared to the number densities $N_x$, $N_y$ of single molecules (monomers) $X$ and $Y$ in the parent phase (consisting of molecules $X$ and $Y$), $N_{xy} \ll N_x, N_y$, the pre-exponential factor $F$ of the equilibrium size distribution function in Eq. (10) takes the form

$$F = N_x + N_y. \tag{20}$$

Accordingly, in three different situations investigated by Reiss [17], it was assumed that no inert carrier gas was present in the parent phase. As applied to a lattice gas (with a lattice site density $N_0$), this assumption corresponds to the complete filling of the lattice sites with monomers, i.e. $N_x + N_y = N_0$. This approach was a generalization of the Frenkel model [25], which characterizes the size distribution of clusters $X_x$ in a single component solid solution of molecules $X$ in the matrix $Y$ with the number density of nucleation sites $F = N_x$.

The extension of Eq. (20) to the case $N_x + N_y \ll N_0$ was widely criticized in the literature. In particular, Lothe and Pound [22] suggested that degrees of freedom corresponding to the translation of clusters have been neglected in the development of nucleation theory for single component systems. As a result, they predicted that the pre-exponential factor is proportional to the total number density of gas molecules (or lattice sites in the case of a lattice gas) $N_0$ rather than vapour molecules (monomers), leading to a large discrepancy with the previous approach. A similar conclusion as applied to the lattice gas was made in a large number of subsequent works, reviewed and supported in [23] (and also used by Russel [11]).

This disagreement ('translation paradox') was discussed by Reiss and Katz [24], who evaluated the partition function of the system taking into account permutations of monomers among clusters and showed that Lothe and Pound's correction to the nucleation theory does not arise (for unary vapours). However, in their subsequent paper [25], where the main qualitative conclusions of [24] were reaffirmed, a correction factor of several orders of magnitude was calculated (however, much smaller than Lothe and Pound's correction). Presumably for this reason, Katz disregarded his previous results [24] and modified the Frenkel model similarly to Lothe and Pound in his subsequent works (e.g. in [26, 27]).

Therefore, the contradiction between different approaches has not been completely resolved and required further analysis. Such an analysis for unary systems was carried out in the recent work of the author [28] within the framework of the general thermodynamic approach [29], taking into account the interaction of monomers with clusters (considered in the statistical mechanics approach [24, 25] and disregarded in the Lothe and Pound model [22, 23]). The excess (or mixing) entropy calculated thermodynamically in [28] was consistent with the value calculated in the statistical approach by Reiss, Kegel and Katz [30], which confirmed the original conclusion of [24].

In particular, it was shown in [28] (for unary systems) that erroneous prediction, $F = N_0$, of the Lothe and Pound model is associated with considering a mixture of monomers and clusters in the ideal gas approximation, neglecting their interaction; whereas their interaction can be taken into account in the weak solution approximation, which leads to $F = N_x$. A generalization of this consideration to binary gas systems, leading to Eq. (20), was given in the author's paper [31], and is extended to binary solid solutions in the Appendix A.

### 3.2. Nucleation rate

When applying the Reiss theory to the nucleation of incoherent particles, the index $x$ will be assigned to solute atoms and the index $y$ to vacancies in the matrix. Results of calculations of the elements of the matrix $\mathbf{D} = (D_{ij})$ and other related parameters of Eq. (17) are presented in the Appendix B, where it is assumed that $D_v c_v^{(0)} \geq D_x c_x$, taking into account that in the majority of metals, the self-diffusion coefficient $D_s$ is determined by the vacancy mechanism and thus $D_s \approx D_v c_v^{(0)}$ [12], and that $D_x \approx D_s$ for Si in Al [32], whereas the typical atomic concentration of Si in Al in the precipitation tests [5–7] was $c_x < 0.01$. In particular, it is confirmed that $\det \mathbf{D} < 0$ (i.e. the critical point is a saddle), and thus

$$(-\det \mathbf{D})^{\frac{1}{2}} = \frac{1}{x^*} \frac{4\mu\Omega}{3} \left(\frac{1}{1+\varphi}\right)^{\frac{4}{3}} \left[\frac{kT}{4\mu\Omega}\left(\ln S_x + \varphi \ln S_v + \frac{3}{8}\frac{kT}{\mu\Omega}\left(\ln\frac{S_x}{S_v}\right)^2\right)\right]^{\frac{1}{2}}, \tag{21}$$

and that

$$D'_{11} \approx D_{11} \approx -\frac{1}{3}\left(\frac{1}{1+\varphi}\right)^{\frac{2}{3}} \frac{1}{x^*} kT \left[\ln S_x + \varphi \ln S_v + \frac{3}{8}\frac{kT}{\mu\Omega}\left(\ln\frac{S_x}{S_v}\right)^2\right], \tag{22}$$

is negative above the critical supersaturation $S_x^*$ and thus provides a maximum of the free energy at the critical point in the passage direction ($x'$-axis) with

$$\tan\theta \approx \varphi - \frac{3kT}{4\mu\Omega}\ln\frac{S_x}{S_v} \ll 1. \tag{23}$$

Therefore, in the case of thermal vacancies in the matrix (with $c_v^{(0)} \ll c_x$), the nucleation rate of incoherent particles (number per unit volume per unit time) takes the form

$$\dot{N} \approx 2\pi D_x c_x^2 \frac{\gamma}{kT}\left(\frac{kT}{\mu\Omega}\right)^{\frac{1}{2}} \ln^{-\frac{1}{2}} S_x \exp\left(-\frac{16\pi\gamma^3\Omega^2}{3(kT)^3 \ln^2 S_x}\right). \tag{24}$$

whereas in the case of quenched samples with an excess of vacancies in the matrix, a more general expression is derived,

$$\dot{N} \approx 2\pi \frac{\gamma}{kT}\left(\frac{kT}{\mu\Omega}\right)^{\frac{1}{2}} \frac{D_x c_x (c_x + c_v)}{\left[\ln S_x + \varphi \ln S_v + \frac{3kT}{8\mu\Omega}\left(\ln\frac{S_x}{S_v}\right)^2\right]^{\frac{1}{2}}} \exp\left\{-\frac{16\pi\gamma^3\Omega^2}{3(kT)^3\left[\ln S_x + \varphi \ln S_v + \frac{3kT}{8\mu\Omega}\left(\ln\frac{S_x}{S_v}\right)^2\right]^2}\right\}. \tag{25}$$

In particular, it may be concluded that an excess of vacancies in quenched samples not only reduces the nucleation barrier, but also increases the pre-exponential factor due to increased diffusion of dissolved atoms, $D_x \propto c_v$, in materials with a vacancy diffusion mechanism.

However, the form of pre-exponential factor in Eq. (25) is essentially different from that obtained in [11], where the unary nucleation theory was used. The reason for such a contradiction was discussed in the author's paper [31], where it was shown that Reiss' expression for the binary nucleation rate, Eq. (17), is valid if $|\det \mathbf{D}|/|D'_{11}| \ll \pi kT$, which corresponds to $\frac{1}{8\pi^2}\frac{\mu\Omega}{kT}\left(kT/\gamma\Omega^{\frac{2}{3}}\right)^3 (\ln S_x)^3 \approx \left(\frac{\ln S_x}{20}\right)^3 \ll 1$, or $\ln S_x \ll 20$, whereas the expression for the unary nucleation rate becomes valid in the opposite limit, $|\det \mathbf{D}|/|D'_{11}| \gg \pi kT$, or $\ln S_x \gg 20$. In the unary limit, corresponding to the condition of a narrow saddle point passage width (so-called 'quasi-classical approximation'), when only one ('classical') trajectory (passing through the critical point $(x^*, y^*)$) gives contribution to the integral in Eq. (16), the nucleation rate reduces to

$$\dot{N}_u \approx J(0) = F\frac{\beta_x^* \beta_y^*(1+\tan^2\theta)}{\beta_y^* + \beta_x^* \tan^2\theta}\left(\frac{|D'_{11}|}{\pi kT}\right)^{1/2} \exp\left(-\frac{\Delta G_0^*}{kT}\right) \approx F\beta_x^* Z \exp\left(-\frac{\Delta G_0^*}{kT}\right), \tag{26}$$

where $Z = \left(\frac{|D'_{11}|}{\pi kT}\right)^{1/2} \approx \left(\frac{|D_{11}|}{\pi kT}\right)^{1/2} = \left(-\frac{1}{2\pi kT}\frac{\partial^2 \Delta G_0(x^*,y^*)}{\partial x^2}\right)^{1/2}$ consistently converges to the Zeldovich factor in the classical (unary) nucleation theory [15].

As a result, Eq. (26) converges to the expression for unary nucleation,

$$\dot{N}_u \approx \left(2^{\frac{1}{2}}/\Omega^{\frac{2}{3}}\right) D_x c_x (c_x + c_v) \left(kT/\gamma\Omega^{\frac{2}{3}}\right)^{\frac{1}{2}} \left[\ln S_x + \varphi \ln S_v + \frac{3}{8}\frac{kT}{\mu\Omega}\left(\ln\frac{S_x}{S_v}\right)^2\right] \exp\left(-\frac{\Delta G_0^*}{kT}\right), \quad (27)$$

which differs from the binary nucleation rate calculated from Eq. (25),

$$\frac{\dot{N}_u}{\dot{N}} = \frac{1}{2^{\frac{1}{2}}\pi}\left(\frac{\mu\Omega}{kT}\right)^{\frac{1}{2}}\left(kT/\gamma\Omega^{\frac{2}{3}}\right)^{\frac{3}{2}}\left[\ln S_x + \varphi \ln S_v + \frac{3}{8}\frac{kT}{\mu\Omega}\left(\ln\frac{S_x}{S_v}\right)^2\right]^{\frac{3}{2}} \approx 0.04 \left[\ln S_x + \varphi \ln S_v + \frac{3}{8}\frac{kT}{\mu\Omega}\left(\ln\frac{S_x}{S_v}\right)^2\right]^{\frac{3}{2}}. \quad (28)$$

In the tests [5, 6] with typical supersaturation $S_x \leq 10^2$, or $\ln S_x \leq 4.5$, the unary approximation is inapplicable and Eq. (27) leads to a significant underestimation of the nucleation rate, according to Eq. (28), especially in the small vicinity of the critical supersaturation, $\ln S_x^* \approx -\varphi \ln S_v$. In Russel's model [11], this underestimation partially compensates for the strong overestimation of the density of nucleation sites $F$ by $(c_x + c_v)^{-1}$ times, as explained in Section 3.1, which reaches several orders of magnitude under test conditions. Therefore, an adequate analysis of the tests can only be carried out within the framework of the binary nucleation theory, using Eqs (24) and (25).

The new model can be used to interpret precipitation kinetics after various thermal treatments (e.g. studied in tests [4–8]) by implementing it in numerical algorithms that treat nucleation, growth and coarsening as coupled processes. An example of such an algorithm is presented in [33], where, based on the classical nucleation theory for unary systems, a numerical code was developed to simulate the evolution of the particle size distribution function during non-isothermal transformations. A similar numerical code was later used in [8] to analyse phase transformations in Al-Si alloys observed in their tests. The model for binary nucleation kinetics proposed in the present work avoids the shortcomings of the simplified unary nucleation theory discussed above and thereby can help improve the predictions of numerical codes based on this theory.

## 4. Conclusion

The traditional theory of homogeneous nucleation of precipitates in solid solutions [1–3] is critically analysed. It is demonstrated that the theory's prediction concerning the increase in the nucleation free energy due to elastic strain, caused by the difference in atomic volumes of the two phases, is applicable to coherent precipitates, but becomes incorrect when applied to incoherent precipitates. Specifically, taking into account thermal point defects in the parent phase, which can be absorbed at the particle-matrix interface during particle nucleation, leads to relaxation of the nuclei, elimination of the contribution of elastic strain energy to the nucleation barrier, and restoration of critical supersaturation $S_x^* \approx 1$ (overestimated in the traditional approach).

On this basis, within the framework of the Reiss theory for binary homogeneous nucleation, a kinetic model is developed to calculate the rate of nucleation of incoherent precipitates in a supersaturated single component solid solution, taking into account the absorption of thermal vacancies at the particle-matrix interface.

The scope of the model is expanded to include excess vacancies arising under non-equilibrium conditions encountered during quenching tests of dilute alloys (e.g. Al-Si). It is confirmed that an excess of vacancies in the quenched samples lowers the nucleation barrier and shifts the critical oversaturation to the value $S_x^* \approx -\varphi \ln S_v$, as was first shown by Russel [11]; however, the pre-exponential kinetic factor of the nucleation rate calculated in the new approach may differ from the predictions of the simplified model [11], developed within the framework of the traditional (unary) nucleation theory, by several orders of magnitude.

Therefore, the new model for binary nucleation kinetics proposed in the present work avoids the shortcomings of the simplified theory of unary nucleation and thereby can help improve the predictions of numerical codes [33, 8] that treat nucleation, growth and coarsening as coupled processes, but the analysis of nucleation kinetics is based on the unary theory.

**Acknowledgements**

The author thanks Dr. V. Tarasov (IBRAE, Moscow) for careful reading of the manuscript and valuable recommendations.

**References**

1. J.W. Christian, The theory of transformations in metals and alloys. Pergamon, Oxford, 1975.
2. R.W. Balluffi, S.M. Allen and W.C. Carter, Kinetics of materials. John Wiley & Sons, 2005.
3. R.E. Smallman, A.H.W. Ngan, Physical metallurgy and advanced materials. 7th ed. Elsevier, 2007.
4. D. Turnbull, H.S. Rosenbaum and H.N. Treaftis, Kinetics of clustering in some aluminium alloys. Acta Metallurgica 8 (1960), pp. 277–295.
5. H.S Rosenbaum and D. Turnbull, On the precipitation of silicon out of a supersaturated aluminum-silicon solid solution. Acta Metallurgica 6 (1958), pp. 653–659.
6. H.S. Rosenbaum and D Turnbull, Metallographic investigation of precipitation of silicon from aluminum. Acta Metallurgica 7 (1959), pp. 664–674.
7. E. Ozawa and H. Kimura, Excess vacancies and the nucleation of precipitates in aluminum-silicon alloys. Acta Metallurgica 18 (1970), pp. 995–1004.
8. F. Lasagni, B. Mingler, M. Dumont, and H.P. Degischer, Precipitation kinetics of Si in aluminium alloys. Materials Science and Engineering A, 480 (2008), pp.383–391.
9. E. Hornbogen and E. A. Starke Jr., Overview no. 102 Theory assisted design of high strength low alloy aluminum. Acta Metallurgica et Materialia 41 (1993), pp. 1–16.
10. F.R.N. Nabarro, The influence of elastic strain on the shape of particles segregating in an alloy. Proc. Phys. Soc. 52 (1940), pp. 90–104.
11. K.C. Russel, The role of excess vacancies in precipitation. Scripta Metallurgica 3 (1969), pp. 313–316.
12. H. Mehrer, Diffusion in solids: fundamentals, methods, materials, diffusion-controlled processes. Springer Series in Solid State Science, Vol. 155, Springer (2007).
13. M. Volmer and A. Weber, Keimbildung in übersättigten Gebilden. Z. Phys. Chem. 119 (1926), pp. 277–303.
14. R. Becker and W. Doering, Kinetische Behandlung der Keimbildung in übersättigten Dämpfen. Ann. Phys. 24 (1935), pp. 719–752.
15. Ja.B. Zeldovich, On the theory of new phase formation: cavitation. Acta Physicochim. URSS 18 (1943), p. 1.
16. H. Reiss, The kinetics of phase transitions in binary systems. J. Chem. Phys. 18 (1950), pp. 840–848.
17. A. Katok, B. Hasselblatt, Introduction to the Modern Theory of Dynamical Systems. Cambridge: Cambridge University Press (1995).

18. J.S. Langer, Statistical theory of the decay of metastable states. Annals of Physics 54 (1969), pp. 258–275.
19. D. Stauffer, Kinetic theory of two-component ("hetero-molecular") nucleation and condensation. J. Aerosol Sci. 7 (1976), pp. 319–333.
20. L.M. Berezhkovskii and V.Yu. Zitserman, Direction of the nucleation current through the saddle point in the binary nucleation theory and the saddle point avoidance. J. Chem. Phys. 102 (1995), pp. 3331–3336.
21. J. Frenkel, Kinetic Theory of Liquids. Dover Publication, New York, 1955.
22. J. Lothe and G.M. Pound, Reconsiderations of nucleation theory. J. Chem. Phys. 36 (1962), pp. 2080–2085.
23. D. Kashchiev, Nucleation: Basic Theory with Applications. Butterworth Heinemann, Oxford, Boston, 2000.
24. H. Reiss and J.L. Katz, Resolution of the translation – rotation paradox in the theory of irreversible condensation. J. Chem. Phys. 46 (1967), pp. 2496–2499.
25. H. Reiss, J.L. Katz and E.R. Cohen, Translation–rotation paradox in the theory of nucleation. J. Chem. Phys. 48 (1968), pp. 5553–5560.
26. J.L. Katz and H. Wiedersich, Nucleation of voids in materials supersaturated with vacancies and interstitials, J. Chem. Phys. 55 (1971), pp. 1414–1425.
27. J.L. Katz, Homogeneous nucleation theory and experiment: A survey. Pure&Appl. Chem, 64 (1992), pp. 1661–1666.
28. M.S. Veshchunov, On the theory of void nucleation in irradiated crystals. J. Nucl. Mater. 571 (2022) 154021.
29. L.D. Landau and E.M. Lifshitz, Theoretical Physics, Vol. 5: Statistical Physics. Pergamon Press, 1980.
30. H. Reiss, W.K. Kegel and J.L. Katz, Resolution of the problems of replacement free energy, 1/S, and internal consistency in nucleation theory by consideration of the length scale for mixing entropy. Phys. Rev. Lett. 78 (1997), pp. 4506–4509.
31. M.S. Veshchunov, Development of the Reiss theory for binary homogeneous nucleation of aerosols. Aerosol Sci. Technol. 58 (2023), pp. 1–10.
32. S.-I. Fujikawa, , K.-I. Hirano and Y. Fukushima, Diffusion of silicon in aluminium. Metallurgical Transactions A 9 (1978), pp. 1811–1815.
33. O.R. Myhr, and Ø. Grong, Modelling of non-isothermal transformations in alloys containing a particle distribution. Acta Materialia 48 (2000), pp.1605–1615.

**Appendix A. Calculation of the pre-exponential factor of the equilibrium size distribution function**

Although the Lothe and Pound approach [22] correctly identified the limitations of the earlier approach (in which the presence of an inert carrier gas was ignored), it inherited the main drawback of this approach, considering the system of monomers and clusters as an ideal mixture.

Indeed, such consideration is valid only in the case of Boltzmann statistics (to which the ideal gas obeys), when all particles are distributed over different thermodynamic states completely independently of each other [29]. For clusters of finite sizes, their interaction with monomers (described in the statistical mechanics approach [24, 25] by permutations of monomers among clusters), cannot be neglected, since clusters, in contrast to monomers, cannot be considered as point particles.

In accordance with general thermodynamics, the additivity of thermodynamic quantities, such as free energy or entropy, is preserved only as long as the interaction between different parts of the system is negligible, as in the case of ideal gas mixtures, for which, for example, the entropy of the mixture is equal to the sum of the entropies of each of gases. Therefore, for a non-ideal mixture of several substances (for example, monomers and clusters), the entropy is no longer equal to the sum of the entropies of each of the substances [29].

To find the excess entropy of a mixture of monomers and clusters, let $\Phi_0(P, T, N_x, N_y)$ be the Gibbs free energy of an ideal solid solution in the crystal matrix (with the number density of lattice sites $N_0$) of monomers $X$ and $Y$ (with the number density $N_x$ and $N_y$, respectively), whose chemical potentials are $\mu_i(P, T, c_i) = \psi_i(P, T) + kT \ln c_i$, where $i = x, y$, and $c_i = N_i/N_0 \ll 1$. Let $\alpha_{xy}$ denote the small change which would occur in the free energy if one spherical cluster $X_x Y_y$ (a nucleus of the new phase) consisting of $(x, y)$ monomers was added to the system. In the thermodynamic approach, clusters are considered as 'macroscopic' subsystems (or 'bodies') with $n_x, n_y \gg 1$, homogeneously distributed in the 'external medium' (solid solution of monomers $X$ and $Y$).

Due to the interactions of clusters with monomers, $X_x Y_y \pm X = X_{x \pm 1} Y_y$, and $X_x Y_y \pm Y = X_x Y_{y \pm 1}$, $\alpha_{xy}$ should be sought as a function of $N_x$ and $N_y$, i.e. $\alpha_{xy} = \alpha_{xy}(P, T, N_x, N_y)$. Due to $N_{xy} \ll N_x, N_y$, where $N_{xy}$ is the number (per unit volume) of clusters of size $(x, y)$, interactions between clusters can be neglected, and thus the free energy takes the form

$$\Phi = N_x \mu_x + N_y \mu_y + N_{xy} \alpha_{xy}(P, T, N_x, N_y) + kT \ln(N_{xy}!), \qquad (A.1)$$

where the translational entropy term, $kT \ln(N_{xy}!) \approx kT N_{xy} \ln(N_{xy}/e)$, takes into account that all (spherical) clusters of one size $(x, y)$ are identical and, being 'macroscopic bodies', are homogeneously distributed in the 'external medium'. Accordingly, Eq. (A.1) can be represented in the form

$$\Phi = N_x \mu_x + N_y \mu_y + kT N_{xy} \ln\left[\frac{N_{xy}}{e} \exp\left(\frac{\alpha_{xy}}{kT}\right)\right]. \qquad (A.2)$$

This consideration is principally different from the Lothe and Pound approach [22], where clusters were considered as a new ideal lattice gas $Z$ with density $N_z \ll N \approx N_0$ added to the existing lattice gas mixture of monomers $X$ and $Y$, and, therefore, become a constituent part of the 'external medium'; this transforms the configurational entropy, $kT \ln\left(\frac{N!}{N_0! N_x! N_y!}\right) \approx -kT \left[N_0 \ln\left(\frac{N_0}{N}\right) + N_x \ln\left(\frac{N_x}{N}\right) + N_y \ln\left(\frac{N_y}{N}\right)\right] \approx -kT \left[N_x \ln\left(\frac{N_x}{N}\right) + N_y \ln\left(\frac{N_y}{N}\right)\right]$ (which enters $\Phi$ through the chemical potential terms), into $kT \ln\left(\frac{(N+N_z)!}{N_0! N_x! N_y! N_z!}\right)$, and hence the additional entropy term in Eq. (A.1) would be $kT N_z \ln\left(\frac{N_z}{N}\right)$, instead of $kT N_z \ln(N_z/e)$, with simultaneous vanishing of the interaction term $N_z \alpha_z$.

Since $\Phi$ in Eq. (A.2) must be a homogeneous function of the first order in $N_x$, $N_y$ and $N_{xy}$ [29], the term $\exp[\alpha_{xy}(P, T, N_x, N_y)/kT]$ in the argument of the logarithm should be sought in the most general form $f_{xy}(P, T)/(N_x + \beta N_y)$. Given that after redefining $x \leftrightarrow y$, the free energy should not change, we can conclude that $\beta = 1$. Accordingly,

$$\Phi = N_x \mu_x + N_y \mu_y + kT N_{xy} \ln\left[\frac{N_{xy}}{e(N_x + N_y)} f_{xy}(P, T)\right], \qquad (A.3)$$

or, introducing a new function $\psi_{xy}(P, T) = kT \ln f_{xy}(P, T)$,

$$\Phi = N_x \mu_x + N_y \mu_y + N_{xy} \psi_{xy}(P, T) + kT N_{xy} \ln\left[\frac{N_{xy}}{e(N_x + N_y)}\right]. \qquad (A.4)$$

Comparison of Eq. (A.4) with Eq. (A.1) shows that

$$N_{xy}\alpha_{xy}(P,T,N_x,N_y) = N_{xy}\psi_{xy}(P,T) - kTN_{xy}\ln(N_x + N_y). \tag{A.5}$$

Therefore, since the first term in Eq. (A.5), $N_{xy}\psi_{xy}(P,T)$, does not depend on the number of monomers, the value $\psi_{xy}(P,T)$ is the standard free energy of a cluster, while the second term of Eq. (A.5), $kTN_{xy}\ln(N_x + N_y)$, is the excess entropy of the mixture.

This leads to the following expressions for the chemical potentials of the 'solvents'

$$\mu'_x = \frac{\partial \Phi}{\partial N_x} = \mu_x - kTc_{xy} \approx \mu_x, \tag{A.6}$$

$$\mu'_y = \frac{\partial \Phi}{\partial N_y} = \mu_y - kTc_{xy} \approx \mu_y, \tag{A.7}$$

where $c_{xy} \approx N_{xy}/(N_x + N_y) \ll 1$, and of the 'solute'

$$\mu_{xy} = \frac{\partial \Phi}{\partial N_{xy}} = kT\ln c_{xy} + \psi_{xy}. \tag{A.8}$$

Therefore, from the equilibrium condition of the chemical reaction $xX + yY = X_xY_y$,

$$x\mu_x + y\mu_y = \mu_{xy}, \tag{A.9}$$

the mass action law can be derived as

$$c_{xy} \approx N_{xy}/(N_x + N_y) = K_{xy}(T), \tag{A.10}$$

with the equilibrium constant

$$K_{xy}(T) = \exp\left(-\frac{\Delta G_0(x,y)}{kT}\right), \tag{A.11}$$

where $\Delta G_0(x,y) = \psi_{xy} - x\mu_x - y\mu_y$ is the Gibbs free energy of formation of a cluster.

If concentrations of clusters of other sizes are also small, their contributions to the total free energy of the system are linear; therefore, the equilibrium size distribution function has the form

$$f_0(x,y) = (N_x + N_y)\exp(-\Delta G_0(x,y)/kT), \tag{A.12}$$

which is derived, as mentioned above, in the thermodynamic approach for 'macroscopic' clusters with $x, y \gg 1$. For this reason, the assertion in Ref. [27] that this expression for a cluster size of 1 does not return the number of monomers is irrelevant.

It is straightforward to see that, considering (following Lothe and Pound [22]) clusters as an ideal lattice gas $Z$ with the chemical potential $\mu_z = \psi_z(P,T) + kTN_z\ln\left(\frac{N_z}{N}\right)$ (as discussed above), the solution to Eq. (A.9) will have the form $c_z = N_z/N_0 = \exp(-\Delta G_0(x,y)/kT)$, where $\Delta G_0(x,y) = \psi_z - x\mu_x - y\mu_y$, and thus the pre-exponential factor in Eq. (A.13) will be equal to the number density of lattice sites, $C = N_0$, derived (erroneously) in [22].

**Appendix B. Calculation of the nucleation rate parameters**

The first and the second derivatives of the free energy, Eq. (9), are calculated as

$$\frac{\partial \Delta G_0(x,n)}{\partial n} = -kT\ln S_v + \frac{8}{3}\pi\gamma\left(\frac{3}{4\pi}\Omega_m\right)^{\frac{2}{3}}x^{-\frac{1}{3}}\left(1+\frac{n}{x}\right)^{-\frac{1}{3}} - \frac{4\mu\Omega}{3}\left(\frac{1}{1+\varphi}\right)^2\left(\varphi - \frac{n}{x}\right), \tag{B.1}$$

$$\frac{\partial \Delta G_0(x,n)}{\partial x} = -kT \ln S_x + \frac{8}{3}\pi\gamma \left(\frac{3}{4\pi}\Omega_m\right)^{\frac{2}{3}} x^{-\frac{1}{3}}\left(1+\frac{n}{x}\right)^{-\frac{1}{3}} + \frac{2\mu\Omega}{3}\left(\frac{1}{1+\varphi}\right)^2 \left[\varphi^2 - \left(\frac{n}{x}\right)^2\right], \tag{B.2}$$

$$\frac{\partial^2 \Delta G_0(x,n)}{\partial n^2} = -\frac{8}{9}\pi\gamma \left(\frac{3}{4\pi}\frac{\Omega}{1+\varphi}\right)^{\frac{2}{3}} x^{-\frac{4}{3}}\left(1+\frac{n}{x}\right)^{-\frac{4}{3}} + \frac{4\mu\Omega}{3}\left(\frac{1}{1+\varphi}\right)^2 \frac{1}{x}, \tag{B.3}$$

$$\frac{\partial^2 \Delta G_0(x,n)}{\partial x^2} = -\frac{8}{9}\pi\gamma \left(\frac{3}{4\pi}\frac{\Omega}{1+\varphi}\right)^{\frac{2}{3}} x^{-\frac{4}{3}}\left(1+\frac{n}{x}\right)^{-\frac{4}{3}} + \frac{4\mu\Omega}{3}\left(\frac{1}{1+\varphi}\right)^2 \frac{1}{x}\left(\frac{n}{x}\right)^2, \tag{B.4}$$

$$\frac{\partial^2 \Delta G_0(x,n)}{\partial n \partial x} = -\frac{8}{9}\pi\gamma \left(\frac{3}{4\pi}\frac{\Omega}{1+\varphi}\right)^{\frac{2}{3}} x^{-\frac{4}{3}}\left(1+\frac{n}{x}\right)^{-\frac{4}{3}} - \frac{4\mu\Omega}{3}\left(\frac{1}{1+\varphi}\right)^2 \frac{1}{x}\frac{n}{x}. \tag{B.5}$$

Accordingly, the elements $D_{ij} = \frac{1}{2}\left.\frac{\partial^2 \Delta G_0(x,y)}{\partial x_i \partial x_j}\right|_{x^*,y^*}$ of the matrix **D** calculated in the first approximation in a small parameter $3kT/4\mu\Omega \sim 10^{-2} \ll 1$ using Eq. (10) take the form

$$D_{11} = \left.\frac{\partial^2 \Delta G_0(x,n)}{\partial x^2}\right|_{x^*,n^*} \approx \frac{4\mu\Omega}{3}\left(\frac{1}{1+\varphi}\right)^2 \frac{1}{x^*}\left[\left(\varphi - \frac{3kT}{4\mu\Omega}\ln\frac{S_x}{S_v}\right)^2 - \frac{kT}{4\mu\Omega}(1+\varphi)^{\frac{4}{3}}\left(\ln S_x + \varphi \ln S_v +\right.\right.$$
$$\left.\left.\frac{3}{8}\frac{kT}{\mu\Omega}\left(\ln\frac{S_x}{S_v}\right)^2\right)\right] \approx -\frac{1}{3}\left(\frac{1}{1+\varphi}\right)^2 \frac{kT}{x^*}(1+\varphi)^{\frac{4}{3}}\left(\ln S_x + \varphi\ln S_v + \frac{3}{8}\frac{kT}{\mu\Omega}\left(\ln\frac{S_x}{S_v}\right)^2\right), \tag{B.6}$$

$$D_{22} = \left.\frac{\partial^2 \Delta G_0(x,n)}{\partial n^2}\right|_{x^*,n^*} \approx \frac{4\mu\Omega}{3}\left(\frac{1}{1+\varphi}\right)^2 \frac{1}{x^*}\left[1 - \frac{\pi kT}{4\mu\Omega}\left(\ln S_x + \varphi\ln S_v + \frac{3}{8}\frac{kT}{\mu\Omega}\left(\ln\frac{S_x}{S_v}\right)^2\right)\right] \approx$$
$$\frac{4\mu\Omega}{3}\left(\frac{1}{1+\varphi}\right)^2 \frac{1}{x^*}, \tag{B.7}$$

$$D_{12} = \left.\frac{\partial^2 \Delta G_0(x,n)}{\partial n \partial x}\right|_{x^*,n^*} \approx -\frac{4\mu\Omega}{3}\left(\frac{1}{1+\varphi}\right)^2 \frac{1}{x^*}\left[\left(\varphi - \frac{3kT}{4\mu\Omega}\ln\frac{S_x}{S_v}\right) + \frac{kT}{4\mu\Omega}(1+\varphi)^{\frac{4}{3}}\left(\ln S_x + \varphi\ln S_v +\right.\right.$$
$$\left.\left.\frac{3}{8}\frac{kT}{\mu\Omega}\left(\ln\frac{S_x}{S_v}\right)^2\right)\right], \tag{B.8}$$

and thus

$$\det \mathbf{D} = D_{11}D_{22} - D_{12}^2 \approx -\left(\frac{1}{x^*}\frac{4\mu\Omega}{3}\right)^2 \left(\frac{1}{1+\varphi}\right)^{\frac{8}{3}} \frac{kT}{4\mu\Omega}\left[\ln S_x + \varphi\ln S_v + \frac{3}{8}\frac{kT}{\mu\Omega}\left(\ln\frac{S_x}{S_v}\right)^2\right], \tag{B.9}$$

which is negative above the critical supersaturation, $\ln S_x > \ln S_x^* \approx -\varphi \ln S_v$, and thus confirms that $(x^*, y^*)$ is a saddle point. This leads to

$$(-\det \mathbf{D})^{\frac{1}{2}} = \frac{1}{x^*}\frac{4\mu\Omega}{3}\left(\frac{1}{1+\varphi}\right)^{\frac{4}{3}}\left[\frac{kT}{4\mu\Omega}\left(\ln S_x + \varphi\ln S_v + \frac{3}{8}\frac{kT}{\mu\Omega}\left(\ln\frac{S_x}{S_v}\right)^2\right)\right]^{\frac{1}{2}}. \tag{B.10}$$

For simplicity, only relatively large values of $|\varphi| \gg \left|\frac{3kT}{4\mu\Omega}\ln\frac{S_x}{S_v}\right| \sim 0.01$, will be further analysed, taking into account that for Si (with $\Omega = a_{Si}^3/8$ and $a_{Si} = 0.5431$ nm) in Al (with $\Omega_m = a_{Al}^3/4$ and $a_{Al} = 0.4049$ nm) $\varphi \approx 0.2$; for Ge (with $\Omega = a_{Ge}^3/8$ and $a_{Ge} = 0.5658$ nm) in Al $\varphi \approx 0.41$; and a negative value $\varphi \approx -0.1$ for incoherent $CuAl_2$ phase in Al. In these cases, $\left|\varphi - \frac{3kT}{4\mu\Omega}\ln\frac{S_x}{S_v}\right| \gg \frac{kT}{4\mu\Omega}(1+\varphi)^{\frac{4}{3}}(\ln S_x + \varphi \ln S_v) \sim 0.01$, and thus Eq. (B.8) can be simplified as

$$D_{12} \approx -\frac{4\mu\Omega}{3}\left(\frac{1}{1+\varphi}\right)^2 \frac{1}{x^*}\left(\varphi - \frac{3kT}{4\mu\Omega}\ln\frac{S_x}{S_v}\right). \tag{B.8'}$$

In the considered case $\beta_x^*/\beta_v^* = D_x c_x / D_v c_v \ll 1$, from Eqs (18) and (19) one obtains

$$\tan\theta \approx \varphi - \frac{3kT}{4\mu\Omega}\ln\frac{S_x}{S_v}, \quad \text{if } D_{12} < 0, \text{ or } \varphi - \frac{3kT}{4\mu\Omega}\ln S_x > 0, \tag{B.11}$$

$$\tan\theta \approx -\left|\varphi - \frac{3kT}{4\mu\Omega}\ln\frac{S_x}{S_v}\right|, \quad \text{if } D_{12} > 0, \text{ or } \varphi - \frac{3kT}{4\mu\Omega}\ln S_x < 0, \tag{B.12}$$

or, more generally,

$$\tan\theta \approx \varphi - \frac{3kT}{4\mu\Omega}\ln\frac{S_x}{S_v} \ll 1, \tag{B.13}$$

where $\tan^2\theta \ll 1$, and

$$\cos^2\theta = \frac{1}{1+\tan^2\theta} \approx \frac{1}{1+\left(\varphi - \frac{3kT}{4\mu\Omega}\ln\frac{S_x}{S_v}\right)^2} \approx 1, \tag{B.14}$$

and thus, taking into account that $\beta_x^* \leq \beta_v^*$ (as explained in Section 3.2),

$$\frac{\beta_x^*\beta_v^*(1+\tan^2\theta)}{\beta_v^* + \beta_x^*\tan^2\theta} \approx \beta_x^*(1+\tan^2\theta) \approx \beta_x^* \approx 8\pi D_x c_x \frac{\gamma}{kT} \frac{1}{\left(\ln S_x + \varphi \ln S_v + \frac{3kT}{8\mu\Omega}\left(\ln\frac{S_x}{S_v}\right)^2\right)}, \tag{B.15}$$

where $\beta_x^* = 4\pi D_x c_x R^*\Omega^{-1}$, and

$$R^* = \left[\frac{3\Omega}{4\pi(1+\varphi)}\right]^{1/3}(x^* + n^*)^{1/3} \approx \left(1 - \frac{kT}{4\mu\Omega}\ln\frac{S_x}{S_v}\right)\frac{2\gamma\Omega}{kT\left[\ln S_x + \varphi \ln S_v + \frac{3kT}{8\mu\Omega}\left(\ln\frac{S_x}{S_v}\right)^2\right]} \approx$$

$$\approx \frac{2\gamma\Omega}{kT\left[\ln S_x + \varphi \ln S_v + \frac{3kT}{8\mu\Omega}\left(\ln\frac{S_x}{S_v}\right)^2\right]}, \tag{B.16}$$

By substituting Eqs (B.13)–(B.15) into Eq. (15), one obtains

$$D'_{11} \approx D_{11} \approx -\frac{1}{3}\left(\frac{1}{1+\varphi}\right)^{\frac{2}{3}}\frac{1}{x^*}kT\left(\ln S_x + \varphi \ln S_v + \frac{3}{8}\frac{kT}{\mu\Omega}\left(\ln\frac{S_x}{S_v}\right)^2\right), \tag{B.17}$$

which is negative above the critical supersaturation, $\ln S_x > \ln S_x^* \approx -\varphi \ln S_v$, and thus ensures a maximum of the free energy at the critical point in the direction of the $x'$-axis and a positive sign of the r.h.s of Eq. (17), leading to

$$\dot{N} \approx \frac{\gamma}{kT}\left(\frac{kT}{4\mu\Omega}\right)^{\frac{1}{2}}\frac{4\pi D_x c_x(c_x + c_v)}{\left[\ln S_x + \varphi \ln S_v + \frac{3kT}{8\mu\Omega}\left(\ln\frac{S_x}{S_v}\right)^2\right]^{\frac{1}{2}}}\exp\left\{-\frac{16\pi\gamma^3\Omega^2}{3(kT)^3\left[\ln S_x + \varphi \ln S_v + \frac{3kT}{8\mu\Omega}\left(\ln\frac{S_x}{S_v}\right)^2\right]^2}\right\}. \tag{B.18}$$

Taking into account that $c_v^{(0)}$ can be generally neglected compared to $c_x$, in the absence of excess vacancies this equation is reduced to

$$\dot{N} \approx 4\pi D_x c_x^2 \frac{\gamma}{kT}\left(\frac{kT}{4\mu\Omega}\right)^{\frac{1}{2}}\ln^{-\frac{1}{2}}S_x \exp\left(-\frac{16\pi\gamma^3\Omega^2}{3(kT)^3\ln^2 S_x}\right). \tag{B.19}$$

It should be noted that the above expression for the cavity volume, $V_m = (x + n)\Omega_m$, used in Eqs (5) and (9), is applicable only in the case $\Omega/\Omega_m < 2$, which corresponds to $\varphi < 1$, while in the case $2 < \Omega/\Omega_m < 3$ and $\varphi < 2$ the correct expression is $V_m = (2x + n)\Omega_m$. Therefore, the transformation strain $\delta(x, n)$ should be recalculated from the expression

$$\frac{V_p - V_m}{V_p} = 1 - \frac{(2x+n)\Omega_m}{x\Omega} = \left(\varphi - 1 - \frac{n}{x}\right)\left(\frac{1}{1+\varphi}\right) = \left(\tilde{\varphi} - \frac{n}{x}\right)\left(\frac{1}{2+\tilde{\varphi}}\right) = (1+\delta)^3 \approx 1 + 3\delta, \tag{B.20}$$

where $\tilde{\varphi} = \varphi - 1 < 1$.